\documentclass[
aps,                    
prd,                    
showpacs,               
nofootinbib,            
twocolumn,             %
showkeys,               %
preprintnumbers,        %
amsmath,               %
amssymb,               %
floatfix]               
{revtex4}               
\usepackage{graphicx}
\usepackage{color}
\usepackage{bm}
\usepackage{multirow}

\begin{document}
\bibliographystyle{apsrev} 

\title{Shockwaves in Supernovae: New Implications on the Diffuse Supernova
Neutrino Background}

\author{S\'ebastien Galais}
\email{galais@ipno.in2p3.fr}
\affiliation{Institut de Physique Nucl\'eaire, F-91406 Orsay cedex, CNRS/IN2P3 and University of Paris-XI, France}

\author{James Kneller} 
\email{kneller@ipno.in2p3.fr}
\affiliation{Institut de Physique Nucl\'eaire, F-91406 Orsay cedex, CNRS/IN2P3 and University of Paris-XI, France}

\author{Cristina Volpe}
\email{volpe@ipno.in2p3.fr}
\affiliation{Institut de Physique Nucl\'eaire, F-91406 Orsay cedex, CNRS/IN2P3 and University of Paris-XI, France}

\author{J\'er\^ome Gava}
\email{gava@ipno.in2p3.fr}
\affiliation{Institut de Physique Nucl\'eaire, F-91406 Orsay cedex, CNRS/IN2P3 and University of Paris-XI, France}

\begin{abstract}
We investigate shock wave effects upon the diffuse supernova neutrino background
using dynamic profiles taken from hydrodynamical simulations and
calculating the neutrino evolution in three flavors with the S-matrix
formalism. We show that the shock wave impact is significant
and introduces modifications of the relic fluxes by about $20 \%$ and of 
the associated event rates at the level of $10-20 \%$. 
Such an effect is important since it is of the same order as the rate variation introduced when different oscillation scenarios (i.e. hierarchy or $\theta_{13}$) are considered.
In addition, due to the shock wave, the rates become less sensitive 
to collective effects, in the inverted hierarchy and when
$\sin^2 2 \theta_{13}$ is between the Chooz limit and $10^{-5}$.
We propose a simplified model to account for shock wave effects in future predictions.
\end{abstract}

\date{\today}

\maketitle

\section{Introduction}
\noindent
Our understanding of neutrino propagation through supernovae has been revolutionized by the demonstration that non-linear effects and the dynamism of the density profile can have significant impact. The neutrino density close to the last scattering surface at the neutrinosphere is so large that it generates significant off-diagonal contributions to the effective potential describing neutrino evolution in matter \cite{Pantaleone:1992eq}. This neutrino-neutrino interaction gives rise to collective behavior regimes  known as synchronization, bipolar oscillations and spectral splits \cite{Samuel:1993uw,Duan:2006an,Hannestad:2006nj,Balantekin:2006tg,Raffelt:2007xt,Dasgupta:2009mg}. 
At the same time,
the evolution of the supernova density profile as the shock wave races through the  mantle, has been found to 
change the adiabaticity of the high-density (H) resonance \cite{Schirato:2002tg,Lunardini:2003eh,Takahashi:2002yj,Fogli:2003dw,Tomas:2004gr,Fogli:2004ff,Choubey:2006aq,Dasgupta:2005wn,Choubey:2007ga,Kneller:2007kg} engendering instances of multiple H resonances \cite{Kneller:2005hf,Dasgupta:2005wn,Kneller:2007kg} which have been shown to lead to phase effects. Thus, the neutrino evolution and the emerging spectra involve an interplay between these two effects that might lead to observable 
consequences for any future Galactic supernova neutrino signal \cite{Gava:2009pj}. 

Very complementary information to the measurement of neutrinos from a single core-collapse supernova
come from the observation of the diffuse supernova neutrino background (DSNB).
Its detection would represent a crucial step forward in our understanding of neutrino properties, of the star formation rate and of the supernova dynamics. The current upper limits are 6.8 $\times$ 10$^3$ $\nu_e$
cm$^{-2}$s$^{-1}$  with 25 MeV $< E_{\nu_e} <50$ MeV (90 $\%$ C.L.)
from LSD \cite{Aglietta:1992yk} and 1.08 $\bar{\nu}_{e}$ cm$^{-2}$s$^{-1}$ with $E_{\bar{\nu}_e} > 19.3$ MeV from Super-Kamiokande \cite{Malek:2002ns}. 
Next-generation neutrino observatories \cite{Autiero:2007zj}, currently under study, should possess the discovery potential to observe 
the DSNB (see e.g. \cite{Ando:2004hc,Lunardini:2005jf,Yuksel:2007mn} and references therein). To be able to disentangle the information encoded by the explosion mechanism, neutrino properties and the star formation rate it is important to 
observe both the $\bar{\nu}_e$ via the scattering on protons and ${\nu}_e$ through neutrino-nucleus interactions
e.g. on $^{40}$Ar \cite{Cocco:2004ac} or $^{12}$C and $^{16}$O \cite{Volpe:2007qx}.
Another strategy to observe the DSNB is to exploit upgraded technologies, such as the addition of Gadolinium 
to water Cherenkov detectors \cite{Beacom:2003nk}, to reach the sensitivity for a discovery with the running Super-Kamiokande detector.  
Finally, relic supernova neutrinos could be searched for though a geological measurement 
\cite{Krauss:1983zn} of the amount of Technetium-97 produced in Molybdenum-98 ore \cite{Haxton:1987bf}, if high-precision solar data and precise neutrino-nucleus cross sections become available \cite{Lazauskas:2009yh}.  

Simultaneously, progress is being made with regard to the
accuracy of the predictions and many calculations have been performed to predict the relic neutrino fluxes and rates
\cite{Krauss:1983zn,Dar:1984aj,Totani:1995dw,Malaney:1996ar,Kaplinghat:1999xi,Fukugita:2002qw,Ando:2004hc,Lunardini:2005jf,Yuksel:2007mn}.
The star formation rate has now been constrained by combining various astrophysical observations \cite{Strigari:2005hu} even at high redshifts \cite{Yuksel:2008cu}, though the local value at $z=0$ retains a factor of 2 uncertainty.    
The calculations of the neutrino propagation in the supernova now include the effects induced by the resonant flavor conversion in matter and recently also the flux modifications induced by the neutrino-neutrino interaction \cite{Chakraborty:2008zp}. 

Our aim in this paper is to show that the shock wave effects upon the 
DSNB are considerable. To this end we follow numerically the neutrino evolution (with and without neutrino self interactions) in a core-collapse supernova using dynamical density profiles from hydrodynamical simulations. We analyze the impact of shock waves both on the relic (anti)neutrino fluxes and on the associated number of events in an observatory on Earth based on different technologies.
Furthermore, we show that the shockwaves reduce the sensitivity of the DSNB to collective effects when $\theta_{13}$ is above sin$^22 \theta_{13}> 10^{-5}$. A simplified model allows us to quantitavely 
account for the shockwave impact, which we use to test the robustness of our results.

The paper is structured as follows. Section II presents the formalism. The results on the relic neutrino fluxes and associated rates are described in Section III. Our simplified model as well as the discussion on the robustness of the results, also in presence of turbulence, are included as well. Section IV is a conclusion.

\section{Theoretical framework}
\noindent
The DSNB flux at Earth, as a function of neutrino energy $E_{\nu}$, is calculated as 
\begin{equation}
F_{\alpha}(E_{\nu})=\int dz\,\left|\frac{dt}{dz}\right|(1+z)\,R_{SN}(z)\frac{dN_{\alpha}(E'_{\nu})}{dE'_{\nu}}
\end{equation} 
where $z$ is the redshift, $E'_{\nu}=(1+z)E_{\nu}$, $R_{SN}$ is the core-collapse supernova rate per unit comoving volume and $dN_{\alpha}/dE_{\nu}$ is the differential spectra
emitted by each supernova. 
The most involved component of the calculations is the determination of the neutrino spectra  
at the supernova, $dN_{\alpha}/dE_{\nu}$. Our calculation proceeds in two steps, the
first to account for the collective effects and the second the dynamic MSW.  

The neutrino wavefunction $\psi(\mathbf{p},\mathbf{r})$ evolves according to 
the Schroedinger-like equation
\begin{equation}
\imath\frac{d\psi}{dt} = [H_0(E_{\nu}) + H_m(\mathbf{r}) + H_{\nu\nu}(\mathbf{p},\mathbf{r})] \psi
\label{E:Schrodinger}
\end{equation}
where $H_0(E_{\nu})$ is the Hamiltonian describing the vacuum oscillations. In the mass basis $H_0$ is 
diagonal and is related to the flavor basis through $H_0^{(f)}=UH_0^{(m)}U^{\dag}$, where $U$ is the 
unitary Maki-Nakagawa-Sakata-Pontecorvo matrix parameterized by three mixing angles -
$\theta_{12}$, $\theta_{13}$, $\theta_{23}$ - and six phases though only one, the CP phase, $\delta$, 
is relevant for oscillations. The second term in the Hamiltonian is the neutrino interaction with matter and is diagonal in the 
flavor basis i.e. $H^{(f)}_m(\mathbf{r})=diag(V_{e}(\mathbf{r}),V_{\mu}(\mathbf{r}),V_{\tau}(\mathbf{r}))$. Only the differences 
between the potentials are relevant. The difference $V_{e}-V_{\mu}$ is the well known $V_{e}-V_{\mu} =\sqrt{2}\,G_F\,N_e(\mathbf{r})$ with $N_e(\mathbf{r})$ the electron density. At tree level $V_{\mu}=V_{\tau}$ in normal matter because there are no charged leptons present other than electrons; the difference $(V_{\tau}-V_{\mu})$ is due to loop corrections and within the 
standard model its ratio with respect to $V_{e}$ is very small, $\mathcal{O}(10^{-5})$ \cite{Botella:1986wy}, however supersymmetric corrections can increase it to the level of $\mathcal{O}(10^{-3})$ \cite{Gava:2009gt}. The final contribution to the Hamiltonian is the neutrino self-interaction term, calculated as in \cite{Gava:2008rp}, using the single-angle approximation. 
It has been shown that 
this approximation reproduces 
both qualitatively and quantitatively the results of multi-angle calculations \cite{Duan:2006an}.
Using this approximation 
$H_{\nu\nu}(\mathbf{p)}$ 
becomes
 \begin{eqnarray}\label{e:4bis}
H_{\nu \nu} &=&{ \sqrt{2} G_F \over {2 \pi R_{\nu}^2}} D(r/ R_{\nu})
\sum_{\alpha} \int
[\rho_{{\nu}_{\underline{\alpha}}} (q') L_{{\nu}_{\underline{\alpha}}}
(q') \nonumber\\
&&\phantom{{ \sqrt{2} G_F \over {2 \pi R_{\nu}^2}} D(r/ R_{\nu})
\sum_{\alpha}}
- \rho_{\bar{{\nu}}_{\underline{\alpha}}}^*(q')
L_{\bar{{\nu}}_{\underline{\alpha}}}(q')] dq'
\end{eqnarray}
with the geometrical factor
 \begin{equation}\label{e:4tris}
D(r/R_{\nu}) = {1 \over 2} \left[1 - \sqrt{1 - \left({R_{\nu} \over
r}\right)^2 }\right]^2
\end{equation}
\noindent
where the radius of the neutrino sphere is $R_{\nu} = 10~$km,
and $L_{{\nu}_{\underline{\alpha}}} (q')$ 
are the neutrino fluxes at the neutrino-sphere.
The sum in Eq.(\ref{e:4bis}) is over the initial neutrino flavor at the neutrino-sphere while the integral is performed over the different neutrino energies. Both neutrinos and anti-neutrinos contribute to the neutrino-neutrino interaction term.
Once the calculations of this step are completed we construct the two evolution operators $S(r_{\star},0)$ and $\bar{S}(r_{\star},0)$ ($r_{\star}=1000$ km) at each neutrino energy.

\begin{figure}[t]
   \includegraphics[scale=0.3,angle=0]{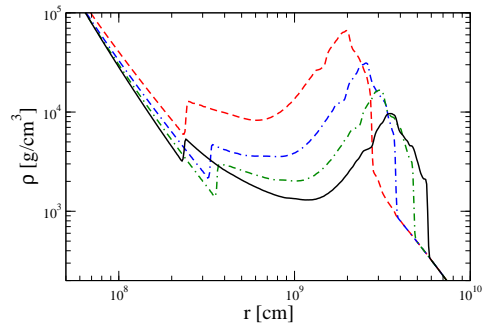}
   \caption{Density profiles with front and reverse shocks as a function of distance in the star. The curves correspond to     1.5 (dashed), 2 (double-dot-dashed), 2.5 (dot-dashed) and 3 seconds (full line) post-bounce.}
   \label{fig:profiles}
\end{figure}
The second step is to determine the (anti)neutrino evolution from $r_{\star}$ 
through to the supernova surface $R$. Here the shock wave effects will appear.
We take supernova matter density profiles from a one-dimensional, hydrodynamical simulation described in 
\cite{Kneller:2007kg} (Figure \ref{fig:profiles}). The explosion of an `initial', standing accretion shock profile is triggered by the injection of $3.36 \times 10^{51}\;{\rm erg}$ of energy into a region beyond a $100\;{\rm km}$ gain radius, in a fashion similar to neutrino heating. The heating creates a wind and falls off exponentially with time. 
The 
profiles contain forward and reverse shocks \cite{Tomas:2004gr,Arcones:2006uq,Kneller:2007kg}, created by the wind, with a large bubble/cavity in-between. The neutrino evolution through the density profiles again follows equation (\ref{E:Schrodinger}) but 
beyond $r\sim 1000\;{\rm km}$
the self interaction effects are negligible. At the end of this step we again construct the evolution operators $S(R,r_{\star})$ and $\bar{S}(R,r_{\star})$.

As in \cite{Gava:2009pj} we suture together the results of the two steps by multiplying in time order the evolution 
operators i.e. $S(R,0)=S(R,r_{\star})\,S(r_{\star},0)$ and $\bar{S}(R,0)=\bar{S}(R,r_{\star})\,\bar{S}(r_{\star},0)$ rather than probabilities 
c.f. \cite{Lunardini:2007vn,Kneller:2007kg,Chakraborty:2008zp}. 
Any possible interference effects are thus captured. We then take into account decoherence due to the divergence of the 
wavepackets and, finally, calculate the $\nu_{e}$ and  $\bar{\nu}_{e}$ survival probabilities $p(E_{\nu},t)$ and 
$\bar{p}(E_{\nu},t)$. 

\begin{figure}[t]
   \includegraphics[scale=0.3,angle=0]{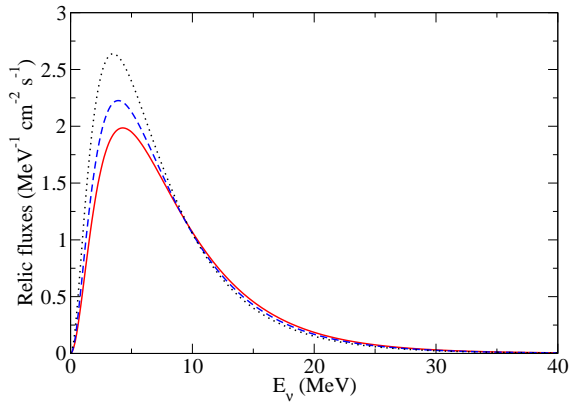}
   \caption{Relic $\nu_e$ flux on Earth, as a function of neutrino
   energy, in the case of normal hierarchy, with the neutrino-neutrino
   interaction and shock wave effects and case L from the numerical
   calculations (dashed line, see text). The other two curves are derived from the analytic formulae \cite{Dasgupta:2007ws,Kneller:2007kg}
for either case S (dotted) or case L (solid) and include the neutrino-neutrino interaction but no shock wave. Note that in the case S the analytical and numerical results are equal.}
\vspace*{3mm}
\label{fig:relicnue}
\end{figure}

The results we present here are obtained with the following parameters. 
At the neutrino sphere radius we adopt Fermi-Dirac distributions for the spectra and make the assumption of energy equipartition among the flavors, $L_{\alpha}=10^{52}\;{\rm erg/ s}$ each at $t=0\;{\rm s}$, with $L(t) \propto \exp(-t/\tau)$, with a cooling time of $\tau=3.5$ s. The average neutrino energies follow a hierarchy i.e. $ \langle E_{\nu_e} \rangle < \langle E_{\bar{\nu}_e} \rangle < \langle E_{\nu_{x}} \rangle$ with values of 12, 15 and 18 MeV respectively. 
We take $\delta m^2_{12}= 8 \times 10^{-5}$eV$^2$, $|\delta m^2_{23}|= 3 \times 10^{-3}$eV$^2$, $\sin^{2} 2\theta_{12}=0.83$ and $\sin^{2} 2\theta_{23}=1$ \cite{Amsler:2008zzb}. The CP phase $\delta$ is here set to zero. For a discussion of
the conditions under which it can modify the neutrino fluxes and its possible effects see \cite{Balantekin:2007es,Gava:2008rp,Kneller:2009vd}.
We shall explore both the normal and inverted hierarchies and consider two cases for the unknown angle $\theta_{13}$: a large value $\sin^{2} 2\theta_{13}=4\times 10^{-4}$ (case L) which lies above above the sin$^22 \theta_{13}> 10^{-5}$ threshold \cite{Dighe:1999bi}, and a much smaller value, $\sin^{2} 2\theta_{13}=4\times 10^{-8}$ (case S), below the threshold. Note that the results corresponding to $\theta_{13}$ large are valid for the range 
sin$^{2} 2 \theta_{13} \in$[0.19,$10^{-5}$]. 

\begin{figure}[t]
   \includegraphics[scale=0.3,angle=0]{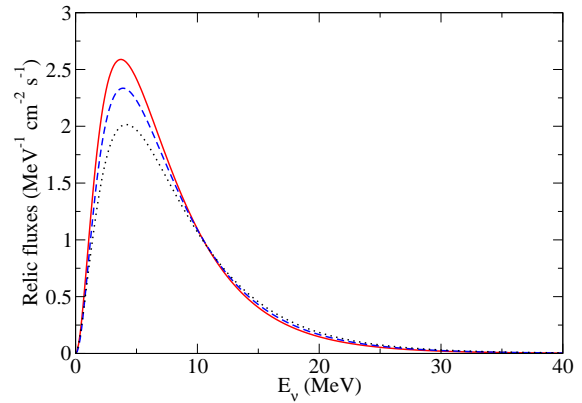}
   \caption{Same as Fig.\ref{fig:relicnue} but for the relic $\bar{\nu}_e$ flux on Earth in the case of inverted hierarchy.}
   \label{fig:relicanue}
\end{figure}
Concerning the other terms that appear in Eq.(\ref{E:Schrodinger}),
we adopt a flat Universe so that
\begin{equation}
dz/dt=-H_{0}(1+z)\sqrt{\Omega_{m}(1+z)^{3}+\Omega_{\Lambda}} 
\end{equation}
and use the concordance $\Lambda CDM$ model parameters $\Omega_{\Lambda}=0.7$ and $\Omega_{m}=0.3$. To calculate $R_{SN}$, we take the star formation rate ($R_{SF}$) from \cite{Yuksel:2008cu}, 
use 
the relation 
\begin{equation}
 {R_{SN}(z)}=R_{SF}(z)
 \frac{{\int_{8M_\odot}^{125M_\odot} \varphi(m)dm}}
 {{\int_{0.1M_\odot}^{125M_\odot} \varphi(m)\ m\ dm}},
\end{equation}
and adopt the initial mass function from \cite{Baldry:2003xi}
\begin{equation}
\varphi(m) \propto \left\lbrace 
 \begin{array}{lc}
  m^{-1.50} & (0.1 M_{\odot} < m < 0.5 M_{\odot})\\ 
  m^{-2.25} & (m > 0.5 M_{\odot})\\ 
 \end{array} \right. .
\end{equation}

\begin{figure}[t]
   \includegraphics[scale=0.29,angle=-90]{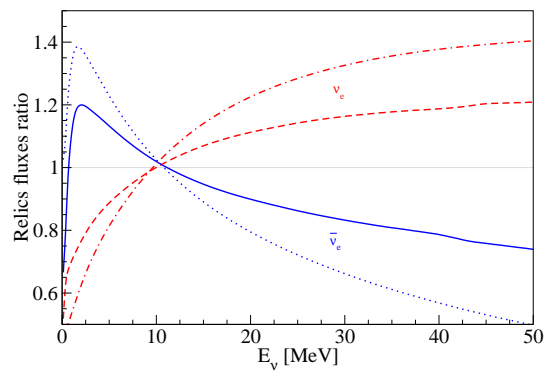}
   \caption{Relic electron (anti-)neutrino flux ratios on Earth, with sin$^{2} 2 \theta_{13}=4\times 10^{-4}$ (case L) over 
sin$^{2}2 \theta_{13}=4\times 10^{-8}$ (case S), as a function of neutrino energy. 
The numerical results concerning the relic fluxes are shown for $\bar{\nu}_e$ in inverted hierarchy (solid) 
and ${\nu}_e$ in normal hierarchy (dashed) include shock wave effects and the $\nu\nu$ interaction. Note that the results corresponding to the larger value of $\theta_{13}$ are valid for the whole range sin$^{2}2 \theta_{13} \in$[0.19,$10^{-5}$]. For comparison, the ratios for $\bar{\nu}_e$ (dotted) and ${\nu}_e$ (dot-dashed) as computed from the analytical formulae \cite{Dasgupta:2007ws,Kneller:2007kg} are also given. In these formulae the
$\nu\nu$ interaction is included but there are no shock wave effects.}   \label{fig:Frelic}
\end{figure}


\section{Results}
\subsection{Results on the DSNB fluxes and rates}
\noindent
Our goal is to show that the shock wave effects upon the DSNB fluxes and on the expected event rates in water Cerenkov, scintillator and liquid argon detectors are significant. 
We then elucidate that the effects can be rather well accounted for by considering four main contributions: the pre-shock, the shock, the phase effects and the post-shock. This quantitative analysis 
allows us to propose a schematic
model to account for shock wave effects in future predictions.

The numerical results for 
the relic fluxes and case L - including the $\nu\nu$ interaction - are shown for ${\nu}_e$ in a normal hierarchy in Figure \ref{fig:relicnue} and for $\bar{\nu}_e$ in the inverted hierarchy in Figure \ref{fig:relicanue}. We also show the relic fluxes computed 
from analytical formulas for comparison where 
we set: $P_{H}=0$ for case L and $P_{H}=1$ for case S for all times \cite{Kneller:2007kg,Chakraborty:2008zp}.
These analytical formulas do not include any shock wave effects (while they include the effect of the neutrino-neutrino interaction).
Figure \ref{fig:Frelic} presents the ratios of the relic (anti)neutrino fluxes for case L value to the case S value so that the 
high energy tail is more visible. 
Note that this ratio is directly sensitive to the shock because the shock affects 
the neutrino propagation only for case L, while for case S
the relic flux is identical to that obtained with the analytical formulas.
Figures \ref{fig:relicnue}-\ref{fig:Frelic} show the analytical results including $\nu\nu$ interaction but no shock wave are at variance with the numerical results which include shock effects : the difference between the analytical fluxes computed with no shocks and numerical ones for the two values of $\theta_{13}$ is reduced roughly by half.


This result is at variance with \cite{Ando:2002zj} where it was found that shock wave effects were negligeable.
Such a difference might be due to the fact that,
as stated in \cite{Schirato:2002tg,Fogli:2003dw,Kneller:2007kg}, the shocks in hydrodynamical profiles are `softened' due to numerical artifacts and are not as steep as they should be. Since the adiabaticity is inversely proportional to the density derivative, a softened shock is more adiabatic than a non-softened shock. This is why the choice for a `large' $\theta_{13}$, as in Ref.\cite{Ando:2002zj}, seems to give an adiabatic result (lower panel of figure 1 in \cite{Ando:2002zj}). 
To compensate for the softness of the shocks in hydro-profiles one must either steepen the shock feature by hand, e.g. as in \cite{Schirato:2002tg,Tomas:2004gr}, or use a value of $\theta_{13}$ (case L) as close as possible to the threshold
sin$^22 \theta_{13}> 10^{-5}$ \emph{but} without making the adiabaticity of the progenitor too small, as in e.g. \cite{Kneller:2007kg}.

\begin{table}[t]
\begin{tabular}{|c|c|c|c|}
\hline
\multicolumn{4}{|c|}{$\bar{\nu}_{e}$ events }\\
\hline
case & Window &   Numerical & Analytical  \\
\hline
~IH~  & 19.3-30 MeV & 0.070 & 0.059\\
\hline
~IH~  & 9.3-25 MeV & 0.190 & 0.176 \\
\hline
~NH~  &  19.3-30 MeV & 0.059 & 0.059  \\
\hline
~NH~  & 9.3-25 MeV & 0.176 & 0.176  \\
\hline
\end{tabular}
\vspace*{5mm}\\
\begin{tabular}{|c|c|c|c|}
\hline
\multicolumn{4}{|c|}{ $\nu_{e}$ events}\\
\hline
case & Window &   Numerical & Analytical \\
\hline
~NH~ & 17.5-41.5 MeV & 0.059 & 0.067\\
\hline
~NH~ &  4.5-41.5 MeV & 0.095 & 0.104\\
\hline
~IH~ & 17.5-41.5 MeV & 0.053  & 0.068 \\
\hline
~IH~ &  4.5-41.5 MeV & 0.089 & 0.105 \\
\hline
\end{tabular}
\caption{Comparison between numerical (with shock wave) and analytical (without shock wave) DSNB event rates (/kton/year),
for the case of large $\theta_{13}$ (L).
The results correspond to $\bar{\nu}_e$+p and
${\nu}_e$+$^{40}$Ar scattering, in different experimental windows, relevant for water Cerenkov, scintillator
and argon detectors. The $\nu\nu$ interaction is included in all cases.}
\label{tab:shock}
\end{table}
\begin{table}[t]
\begin{tabular}{|c|c|c|c|}
\hline
\multicolumn{4}{|c|}{Inverted Hierarchy: with $\nu\nu$ (without $\nu\nu$)}\\
\hline
$N_{\text{events}}$ & Window & L & S \\
\hline
$\bar{\nu}_{e}$ & 19.3-30 MeV & 0.070 (0.070) & 0.080 (0.059)\\
\hline
$\bar{\nu}_{e}$ & 9.3-25 MeV & 0.190 (0.189) & 0.202 (0.176)\\
\hline
\end{tabular}
\vspace*{2mm}\\
\begin{tabular}{|c|c|c|}
\hline
\multicolumn{3}{|c|}{Normal Hierarchy  : with/without  $\nu\nu$}\\
\hline
$N_{\text{events}}$ & Window & L or  S\\
\hline
$\bar{\nu}_{e}$ & 19.3-30 MeV & 0.059 \\
\hline
$\bar{\nu}_{e}$ & 9.3-25 MeV & 0.176 \\
\hline
\end{tabular}
\vspace*{5mm}\\
\begin{tabular}{|c|c|c|c|}
\hline
\multicolumn{4}{|c|}{Normal Hierarchy : with/without  $\nu\nu$}\\
\hline
$N_{\text{events}}$ & Window & L & S\\
\hline
$\nu_{e}$ & 17.5-41.5 MeV& 0.059 & 0.052 \\
\hline
$\nu_{e}$ & 4.5-41.5 MeV& 0.095 & 0.086 \\
\hline
\end{tabular}
\vspace*{1mm}\\
\begin{tabular}{|c|c|c|}
\hline
\multicolumn{3}{|c|}{Inverted Hierarchy: with $\nu\nu$ (without $\nu\nu$)}\\
\hline
$N_{\text{events}}$ & Window & L or S\\
\hline
$\nu_{e}$ & 17.5-41.5 MeV& 0.053 (0.052)\\
\hline
$\nu_{e}$ & 4.5-41.5 MeV& 0.089 (0.086)\\
\hline
\end{tabular}
\caption{DSNB event rates/kton/year for $\bar{\nu}_e$+p and
${\nu}_e$+$^{40}$Ar scattering, from our numerical calculations, relevant for water Cerenkov, scintillator
and argon detectors. The large $\theta_{13}$ (L) and small (S) cases are the same as in Figure \ref{fig:Frelic}.
The calculations include shock waves effects, with/without the
neutrino-neutrino ($\nu\nu$) interaction.}
\label{tab:results}
\end{table}

After computing the relic fluxes we can then compute the DSNB event rates in a detector with both our numerical results 
and using the fluxes computed using the analytic formulae.
We shall consider three types of detector: water Cerenkov, scintillator and  liquid argon. While in the former the main detection channel is $\bar{\nu}_e$+p, in the latter  
${\nu}_e$ are observed through $^{40}$Ar scattering. 
The computed event rates are shown in \ref{tab:shock} and \ref{tab:results}. 

Table \ref{tab:shock} presents a comparison between the event rates for case L obtained with the analytical formulas -- that only take into account a single MSW resonance -- with the numerical calculations which include the shock effects.
Both calculations include the neutrino-neutrino interaction. 
One can see that, indeed, when one includes the shock wave the rates are modified by $10-20 \%$ relative to the analytic formulae showing that the shock wave impact the rates significantly. This is true both in the inverted hierarchy
for $\bar{\nu}_e$ and in the normal hierarchy for ${\nu}_e$. On the other hand, if $\theta_{13}$ is small,  analytical and numerical results are equal in the inverted hierarchy and $\bar{\nu}_e$.  However, for ${\nu}_e$ and normal hierarchy a discrepancy of about 20 $\%$ between the two calculations appear. We have found that such a difference is a combined  effect of  the $V_{\mu\tau}$ refractive index with the $\nu\nu$ interaction. 
Let us now discuss the rate sensitivity to different oscillation scenarios.

Table \ref{tab:results} presents the DSNB rates obtained from the
numerical calculations only for normal/inverted hierarchy and large/small $\theta_{13}$. The DSNB rates change between 7 to 15 $\%$ when one goes from the L to the S case,  both in the neutrino (normal hierarchy) and in the anti-neutrino (inverted  hierarchy) channels.  Such a variation 
is of the same order as the shock wave effects previously discussed.
As already suggested by Figs. \ref{fig:relicnue}-\ref{fig:Frelic}, the
rate variation from L to S cases given by the approximate analytical
formulas is reduced by half compared to the numerical calculations
showing again that including the shock wave is important. Finally the numerical results for $\bar{\nu}_e$ are the same for normal hierarchy if $\theta_{13}$ is L or S.

From Table \ref{tab:results} the effect of the $\nu\nu$ interaction in presence of the shock wave can also be seen. 
If we focus upon $\bar{\nu}_e$ and case L we observe a most surprising result: when shock wave effects are included the $\bar{\nu}_e$ DSNB rates lose their sensitivity to the collective effects. On the other hand, 
the event rates for ${\nu}_e$ in inverted hierarchy increase slightly
because of the spectral split that is induced when the $\nu\nu$
interaction is included. For the $\bar{\nu}_e$ DSNB event and case S,
we see that the $\nu\nu$ interaction alone leads to an increase of around 10-30\%. 

Before going on to explain why the sensitivity to the $\nu\nu$ interaction is lost, in presence of the shock wave, we can use these calculations to predict, on average, 343, 91 and 62 events over 10 years in a detector like MEMPHYS (440 kton), LENA (50 kton) or GLACIER (100 kton) respectively. 

\subsection{A simplified model to account for shock wave effects}
\noindent
To gather further insight on the robustness of our results - and of the loss of sensitivity to the collective effects - we have built up a simplified model for our numerical calculations.
At any given energy the time-integrated $\bar{\nu}_e$ spectra are
\begin{eqnarray}
\frac{dN_{\bar{e}}}{dE} & = & \int_{0}^{\infty}\,dt\,\left( \bar{p} \Phi_{\bar{e}}+\left(1-\bar{p}\right)\Phi_{\bar{x}}\right). \label{dNebar}
\end{eqnarray} 
\begin{figure}[t]
   \includegraphics[scale=0.3,angle=-90]{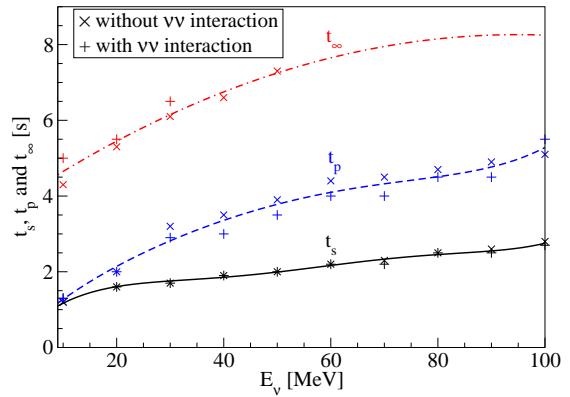}
   \caption{Numerical results on the three times characterizing the neutrino evolution, with shock waves, in the supernova mantle :  the shock $t_s$, the phase-effects $t_p$ and the post-shock $t_{\infty}$. The corresponding analytical fits are also shown.}
   \label{fig:times}
\end{figure}
\noindent
where $\Phi_{\alpha}(E,t)$ are the spectra at the neutrinosphere. The time dependence of $\bar{p}(E,t)$ can be summarized as follows. At around $t \sim 2\;{\rm s}$ into the supernova the shock wave reaches the H resonance region. If the value of $\theta_{13}$ is large then the adiabatic evolution prior to the shock becomes non-adiabatic when the shock arrives. Non-adiabatic propagation persists for some time and then switches to a period where the survival probability oscillates rapidly due to phase effects generated by multiple H resonances. These eventually cease whereupon neutrino propagation enters a post-shock regime that may be adiabatic or semi-adiabatic depending upon $\theta_{13}$ and the post-shock profile. The shock affects lower neutrino energies before the higher and the duration of the non-adiabatic period grows with neutrino energy. This picture allows us to approximate the evolution of $\bar{p}(E,t)$ as a sequence of phases: a) the pre-shock interval up to $t_{s}(E)$, b) the shock interval from $t_{s}(E)$ to $t_{p}(E)$ c) the phase-effect interval from $t_{p}(E)$ to $t_{\infty}(E)$ d) the post-shock period from $t_{\infty}(E)$ onwards. In this way we divide the integral in Eq.\eqref{dNebar} into four. The transition times taken from the numerical results as a function of neutrino energy are shown in Figure \ref{fig:times}. The curves have been fitted with polynomials given by
\begin{eqnarray}\label{e:fit}
 t_{s,p}(E) &=& \sum^{5}_{i=0}{a_{i}.E^{i}}  \\
 t_{\infty}(E) &=& 3.75+9.5\times 10^{-2}\ E-5\times 10^{-4}\ E^{2}
\end{eqnarray}
\begin{table}[!h]
\begin{tabular}{|c|c|c|c|}
  \hline \hline
  & $a_{0}$ & $a_{1}$ & $a_{2}$ \\ 
  \hline
$~~t_{s}~~$   & $~1.02\times 10^{-2}~$ & $~1.72\times 10^{-1}~$ & $~-6.88\times 10^{-3}~$ \\
  \hline
 $~~t_{p}~~$ & $~9.83\times 10^{-2}~$  & $~1.39\times 10^{-1}~$ & $~-2.47\times 10^{-3}~$ \\ \hline
\end{tabular}
\vspace{2mm}\\
\begin{tabular}{|c|c|c|c|}
\hline
\hline
&  $a_{3}$ & $a_{4}$ & $a_{5}$ \\
  \hline
$~~t_{s}~~$ & $~1.4\times 10^{-4}~$  &   $~-1.2\times 10^{-6}~$ & $~4.2\times 10^{-9}~$ \\ \hline
$~~t_{p}~~$ & $~4.0\times 10^{-5}~$ & $~-4.4\times 10^{-7}~$ & $~1.9\times 10^{-9}~$ \\ \hline
\end{tabular} 
\caption{Coefficients for the polynomial fit of Eq.(\ref{e:fit}).}
\label{tab:t-coeff}
\end{table}
\begin{table}[!h]
\begin{tabular}{|c|c|c|c|c|}
  \hline \hline
  Interval & $0\rightarrow t_{s}$ & $t_{s}\rightarrow t_{p}$ & $t_{p}\rightarrow t_{\infty}$ & $t_{\infty}\rightarrow \infty$\\
  \hline
  With $\nu\nu$ & 0.5436 & 0.0634 & 0.3092 & 0.2548 \\
  \hline
  Without $\nu\nu$ & 0.1611 & 0.6356 & 0.3531 & 0.4835 \\
  \hline
\end{tabular} 
\caption{Average $\bar{p}$ values from the numerical calculations, used in the simplified model (see text).}
\label{tab:pbar_values}
\end{table}
with the coefficients given in Table \ref{tab:t-coeff}.
We then approximate $\bar{p}(E,t)$ within each domain by using average
$\bar{p}$ value independent of $E$. These averages are shown in Table \ref{tab:pbar_values}. Using this schematic model we can quantitatively reproduce the rates obtained with or without the
neutrino-neutrino interaction in the inverted hierarchy and any $\theta_{13}$ shown in Tab.\ref{tab:results}.

\begin{figure}
   \includegraphics[scale=0.3,angle=0]{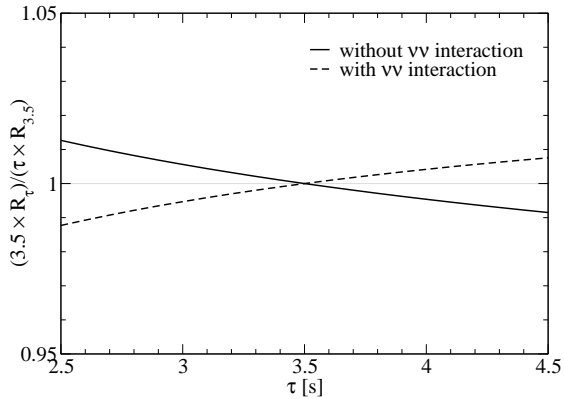}
   \caption{The ratio of the $\bar{\nu}_e$ DNSB event rate with altered cooling time $\tau$ relative to the rate with $\tau=3.5\;{\rm s}$ for the inverted hierarchy and case L either with or without the neutrino self interaction effects. All parameters are as before including the neutrino luminosities so the additional $\tau$ factor accounts for the change in the overall neutrino luminosity.}
   \label{fig:tau}
\end{figure}
\begin{figure}[tbh]
   \includegraphics[scale=0.3,angle=0]{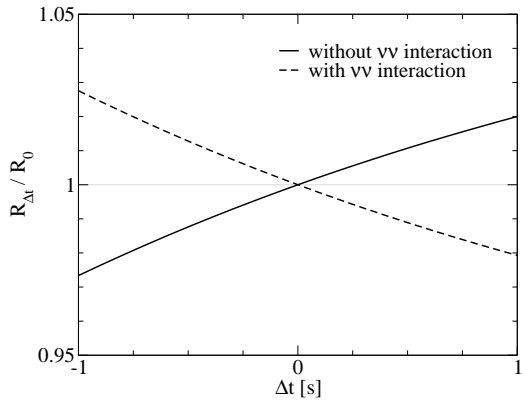}
   \caption{The ratio of the $\bar{\nu}_e$ DNSB event rate when an offset $\Delta t$ time is introduced relative to the rate with zero offset for the inverted hierarchy and case L either with or without the neutrino self interaction effects.}
   \label{fig:deltat}
\end{figure}

We can now use the simplified model just described to understand why, when the shock wave is present, we loose sensitivity
to the collective effects.
Indeed, when computing the DSNB relic fluxes for case L and a given hierarchy the effect of the shock leads to a time-integrated
spectra that is composed of a mixture of pre-shock and post-shock fluxes. Due to the coincidence between the cooling time $\tau = 3.5\;{\rm s}$ we selected and the arrival of the shock in the H resonance region, $t\sim 2\;{\rm s}$ for the hydro model used, this mixture is almost exactly 50:50.
It is because this composition is so close to equality that switching on or off the $\nu\nu$ collective effects has little impact for the $\bar{\nu}_e$ DSNB rates in case L and the inverse hierarchy. But if we change $\tau$ or alter the time at which the shock reaches the H resonances then we alter the mixture of pre- and post-shock fluxes that make the time-integrated spectra and thus might recover some sensitivity to the collective effects. 
To properly examine the sensitivity to the cooling time and/or shock dynamics we can use the simplified model.

\subsection{Robustness of the results}
\noindent
Using the model described in the previous section we have investigated the sensitivity of our results to the neutrino mixing parameters, to the emission spectra and also to the shock dynamics. 
The ratio of the results with an altered $\tau$ relative to the 
fiducial rate at $\tau=3.5\;{\rm s}$ is shown in Figure \ref{fig:deltat}. The additional $\tau$ factor accounts for the change in the integrated luminosity. For cooling times between 2.5 and 4.5 s the DSNB rates 
with and without neutrino-neutrino interaction, differ at most by 2$-$3\%. 
We can also use the model to test the sensitivity to the profile dynamics. By introducing a temporal offset to all the times, so the new times are now $t_{s,p,\infty}+\Delta t$ we can alter the shock arrival time in the H resonance region. The ratio of the rates with non-zero $\Delta t$ relative to the 
rate with no offset are shown in Figure \ref{fig:deltat}. For reasonable values, ($\Delta t \pm 1$s), the difference does not exceed 4\% from the rates listed in Table \ref{tab:results}. Thus we conclude that the loss of sensitivity to $\theta_{13}$ and to the collective effects is robust and is not an effect of the coincidence of our original choice of cooling time and the density profiles adopted.
\begin{figure}[t]
   \includegraphics[scale=0.3,angle=0]{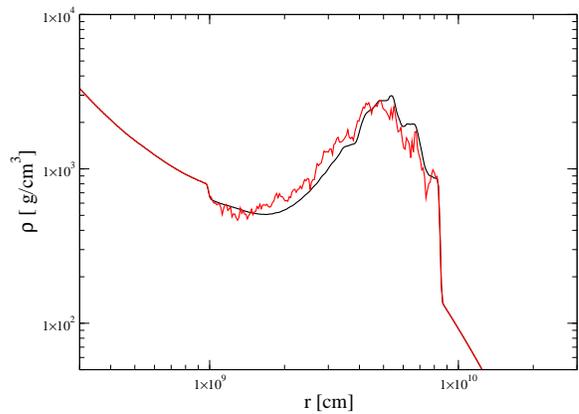}
   \caption{One-dimensional density profile at 4.5 s with (dashed) and without noise (full line, see text).}
   \label{fig:density+noise}
\end{figure}

\begin{figure}[t]
 \includegraphics[scale=0.3,angle=0]{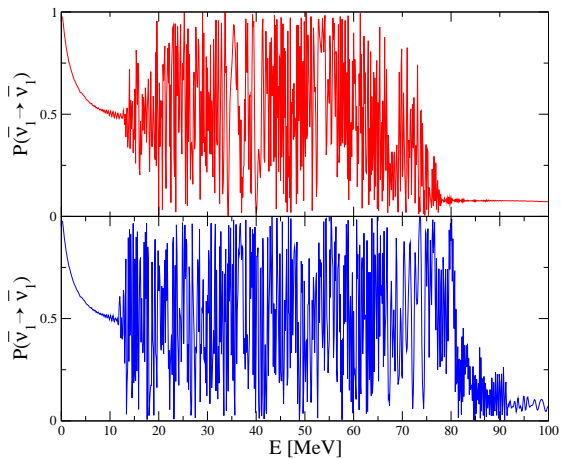}
   \caption{Matter basis survival probability for $\bar{\nu}_1$ for inverted hierarchy and
sin$^2 2 \theta_{13} = 4 \times 10^{-4}$ for the profiles of Figure \ref{fig:density+noise} with no noise (upper) and with noise (lower figure).}
 \label{fig:turbulence}
\vspace*{4mm}
\end{figure}

Let us discuss the robustness of our results with respect to the possible presence of turbulence. 
The transition from adiabatic to non-adiabatic propagation for case L is also a crucial factor in reducing the sensitivity to the collective effects. Any new effect or feature in the profiles that changes the size of the adiabatic to non-adiabatic transition would alter the composition of the time-integrated neutrino spectrum. The presence of turbulence may be such a feature. The restriction to spherical symmetry for the density profiles used in this paper means the profiles do not possess the small scale density fluctuations seen in two dimensional simulations \cite{Tomas:2004gr,Kneller:2007kg} generated by non-radial flow through aspherical shocks. But one also observes that the profiles in two dimensional simulations also possess common features with the one dimensional profiles: they both contain forward and reverse shocks with a cavity between them and any difference are in the details such as the ratio of shock heights or cavity depth. It is because of this similarity that we anticipate that adoption of a turbulent profile would not actually give very different results. Our reasoning is that at any given moment during the neutrino emission, the effect of strong turbulence is to create a range of neutrino energies with survival probabilities that are essentially randomly distributed from zero to unity \cite{Fogli:2006xy,Friedland:2006ta,Choubey:2007ga,Duan:2009cd}. When averaged over an ensemble of fluctuation spectra strong turbulence tends to drive the neutrino oscillation probabilities to one-half.
But, the neutrino oscillation probabilities through our nonturbulent profiles already experience multiple resonances which lead to phases effects and, these too, drive the probabilities averaged over energy and/or time bins to one half. Thus turbulence is only a difference of degree 
rather than of kind from the phase effects already present in our results as long as the turbulence affects the neutrinos during the period when 
they experienced phase effects.

We can investigate this quantitatively by comparing calculations of neutrino propagation through a profile with and without turbulence by adopting the approach of Fogli \emph{et al.} \cite{Fogli:2006xy}. Fluctuations are added to the one dimensional density profile 
at $4.5\;{\rm s}$ by multiplying the potential, $V_{e}(r)$, 
by a factor $1+F(r)$. The function $F(r)$ is assumed to be a 
Gaussian field restricted to the region between the forward shock at $r_s$ and the reverse shock 
at $r_r$. We adopt a Kolmogorov spectrum for the fluctuations with a lower cutoff wavenumber, $k_{\star}$, equal to $k_{\star} = 2\pi/(r_s - r_r)$.
We represent $F(r)$ by a Fourier series with 1000 modes and use the condition that $F(r)$ vanishes at the two shocks. 
Thus $F(r)$ is given by
\begin{equation}
F(r)=\sqrt{C}\,\sum_{n=1}^{1000} A_{n} \sin\left[n\,k_{\star}\,(r-r_r)\right]
\end{equation}
where the $A_n$ are independent random Gaussian variates with 
variance $\sigma^2_n = n^{-5/3}$. The constant $C$ sets the scale for the fluctuations and we
select $C$ to give an expected rms integrated power in the fluctuations of 15\%. 
This satisfies the limit for strong turbulence in the large $\theta_{13}$ case \cite{Friedland:2006ta}.
The density profile with one realization of $F(r)$ is shown in Figure \ref{fig:density+noise} and in Figure \ref{fig:turbulence} we 
compare the antineutrino survival probabilities as a function of energy for the same two profiles. 
We see that turbulence causes an expansion of the neutrino energies experiencing phases effects because the noise 
extends the density range reached between the shocks and thus those neutrinos with resonance densities 
above and below the range with the nonturbulent profile can now be affected. Upon closer inspection we also 
find that the phase effects oscillate with energy more frequently. But despite these quantitative differences between 
the two calculations the overall picture is the same for the two cases leading us to confirm our expectations.

\section{Conclusions}
\noindent
By performing collective and dynamical numerical calculations of relic supernova neutrino fluxes we have found that the shock effects are considerable if $\theta_{13}$ is larger than the threshold of $\sin^{2}2\theta_{13} \gtrsim 10^{-5}$.
These effects are twofold. First 
the shocks reduce the sensitivity to the oscillation parameters ($\theta_{13}$ in particular).
Second there is a loss of sensitivity to the collective effects for the inverse hierarchy and large $\theta_{13}$. 
This loss of sensitivity to the collective effects and $\theta_{13}$ appears to be robust against variations of the neutrino emission, of the details of the density profiles and also against turbulence. 
We have also shown that the use of analytical formulae without the shock effects can significantly overestimate (or underestimate) the rates.
In summary, shock wave effects introduce important modifications of the relic neutrino fluxes and rates and need to be considered in future modelling and simulations\footnote{Note that DSNB calculations based on the 1987 A data as templates for $dN/dE_{\nu}$, e.g. \cite{Lunardini:2005jf} have shock effects implicitly included.}. To assist further efforts to this end we have proposed a simplified procedure to quantitatively capture the relevant shock wave effects, both in the relic neutrino fluxes and the associated event rates in observatories on Earth.



\begin{thebibliography}{99}
\bibitem{Pantaleone:1992eq}
  J.~T.~Pantaleone,
  Phys.\ Lett.\  B {\bf 287}, 128 (1992).

\bibitem{Samuel:1993uw}
  S.~Samuel,
  Phys.\ Rev.\  D {\bf 48}, 1462 (1993).

\bibitem{Duan:2006an}
  H.~Duan, G.~M.~Fuller, J.~Carlson and Y.~Z.~Qian,
  Phys.\ Rev.\  D {\bf 74}, 105014 (2006)
  [arXiv:astro-ph/0606616].

\bibitem{Hannestad:2006nj}
  S.~Hannestad, G.~G.~Raffelt, G.~Sigl and Y.~Y.~Y.~Wong,
  Phys.\ Rev.\  D {\bf 74}, 105010 (2006)
  [Erratum-ibid.\  D {\bf 76}, 029901 (2007)]
  [arXiv:astro-ph/0608695].

\bibitem{Balantekin:2006tg}
  A.~B.~Balantekin and Y.~Pehlivan,
  J.\ Phys.\ G {\bf 34} (2007) 47
  [arXiv:astro-ph/0607527].

\bibitem{Raffelt:2007xt}
  G.~G.~Raffelt and A.~Y.~Smirnov,
  Phys.\ Rev.\  D {\bf 76}, 125008 (2007)
  [arXiv:0709.4641 [hep-ph]].

\bibitem{Dasgupta:2009mg}
  B.~Dasgupta, A.~Dighe, G.~G.~Raffelt and A.~Y.~Smirnov,
  Phys.\ Rev.\ Lett.\  {\bf 103}, 051105 (2009)
  [arXiv:0904.3542 [hep-ph]].

\bibitem{Schirato:2002tg}
  R.~C.~Schirato and G.~M.~Fuller,
  arXiv:astro-ph/0205390.

\bibitem{Lunardini:2003eh}
  C.~Lunardini and A.~Y.~Smirnov,
  JCAP {\bf 0306}, 009 (2003)
  [arXiv:hep-ph/0302033].

\bibitem{Takahashi:2002yj}
  K.~Takahashi, K.~Sato, H.~E.~Dalhed and J.~R.~Wilson,
  Astropart.\ Phys.\  {\bf 20}, 189 (2003)
  [arXiv:astro-ph/0212195].

\bibitem{Fogli:2003dw}
  G.~L.~Fogli, E.~Lisi, D.~Montanino and A.~Mirizzi,
  Phys.\ Rev.\  D {\bf 68}, 033005 (2003)
  [arXiv:hep-ph/0304056].

\bibitem{Tomas:2004gr}
  R.~Tomas, M.~Kachelriess, G.~Raffelt, A.~Dighe, H.~T.~Janka and L.~Scheck,
  JCAP {\bf 0409}, 015 (2004)
  [arXiv:astro-ph/0407132].

\bibitem{Fogli:2004ff}
  G.~L.~Fogli, E.~Lisi, A.~Mirizzi and D.~Montanino,
  JCAP {\bf 0504}, 002 (2005)
  [arXiv:hep-ph/0412046].

\bibitem{Choubey:2006aq}
  S.~Choubey, N.~P.~Harries and G.~G.~Ross,
  Phys.\ Rev.\  D {\bf 74}, 053010 (2006)
  [arXiv:hep-ph/0605255].

\bibitem{Dasgupta:2005wn}
  B.~Dasgupta and A.~Dighe,
  Phys.\ Rev.\  D {\bf 75}, 093002 (2007)
  [arXiv:hep-ph/0510219].

\bibitem{Choubey:2007ga}
  S.~Choubey, N.~P.~Harries and G.~G.~Ross,
  Phys.\ Rev.\  D {\bf 76}, 073013 (2007)
  [arXiv:hep-ph/0703092].

\bibitem{Kneller:2007kg}
  J.~P.~Kneller, G.~C.~McLaughlin and J.~Brockman,
  Phys.\ Rev.\  D {\bf 77}, 045023 (2008)
  [arXiv:0705.3835 [astro-ph]].

\bibitem{Kneller:2005hf}
  J.~P.~Kneller and G.~C.~McLaughlin,
  Phys.\ Rev.\  D {\bf 73}, 056003 (2006)
  [arXiv:hep-ph/0509356].

\bibitem{Gava:2009pj}
  J.~Gava, J.~Kneller, C.~Volpe and G.~C.~McLaughlin,
  Phys.\ Rev.\ Lett.\  {\bf 103}, 071101 (2009)
  [arXiv:0902.0317 [hep-ph]].



\bibitem{Krauss:1983zn}
  L.~M.~Krauss, S.~L.~Glashow and D.~N.~Schramm,
  Nature {\bf 310} (1984) 191.

\bibitem{Dar:1984aj}
  A.~Dar,
  Phys.\ Rev.\ Lett.\  {\bf 55} (1985) 1422.

\bibitem{Totani:1995dw}
  T.~Totani, K.~Sato and Y.~Yoshii,
  Astrophys.\ J.\  {\bf 460} (1996) 303
  [arXiv:astro-ph/9509130].

\bibitem{Malaney:1996ar}
  R.~A.~Malaney,
  Astropart.\ Phys.\  {\bf 7} (1997) 125
  [arXiv:astro-ph/9612012].

\bibitem{Kaplinghat:1999xi}
  M.~Kaplinghat, G.~Steigman and T.~P.~Walker,
  Phys.\ Rev.\  D {\bf 62} (2000) 043001
  [arXiv:astro-ph/9912391].

\bibitem{Fukugita:2002qw}
  M.~Fukugita and M.~Kawasaki,
  Mon.\ Not.\ Roy.\ Astron.\ Soc.\  {\bf 340} (2003) L7
  [arXiv:astro-ph/0204376].

\bibitem{Ando:2004hc}
  S.~Ando and K.~Sato,
  New J.\ Phys.\  {\bf 6}, 170 (2004)
  [arXiv:astro-ph/0410061].

\bibitem{Lunardini:2005jf}
  C.~Lunardini,
  Astropart.\ Phys.\  {\bf 26}, 190 (2006)
  [arXiv:astro-ph/0509233].

\bibitem{Yuksel:2007mn}
  H.~Yuksel and J.~F.~Beacom,
  Phys.\ Rev.\  D {\bf 76}, 083007 (2007)
  [arXiv:astro-ph/0702613].

\bibitem{Aglietta:1992yk}
  M.~Aglietta {\it et al.},
  Astropart.\ Phys.\  {\bf 1}, 1 (1992).
  
\bibitem{Malek:2002ns}
  M.~Malek {\it et al.}  [Super-Kamiokande Collaboration],
  Phys.\ Rev.\ Lett.\  {\bf 90} (2003) 061101
  [arXiv:hep-ex/0209028].

\bibitem{Autiero:2007zj}
  D.~Autiero {\it et al.},
  JCAP {\bf 0711}, 011 (2007)
  [arXiv:0705.0116 [hep-ph]].

\bibitem{Cocco:2004ac}
  A.~G.~Cocco, A.~Ereditato, G.~Fiorillo, G.~Mangano and V.~Pettorino,
  JCAP {\bf 0412}, 002 (2004)
  [arXiv:hep-ph/0408031].

\bibitem{Volpe:2007qx}
  C.~Volpe and J.~Welzel,
  arXiv:0711.3237 [astro-ph].

\bibitem{Beacom:2003nk}
  J.~F.~Beacom and M.~R.~Vagins,
  Phys.\ Rev.\ Lett.\  {\bf 93}, 171101 (2004)
  [arXiv:hep-ph/0309300].

\bibitem{Strigari:2005hu}
  L.~E.~Strigari, J.~F.~Beacom, T.~P.~Walker and P.~Zhang,
  JCAP {\bf 0504} (2005) 017
  [arXiv:astro-ph/0502150].

\cite{Yuksel:2008cu}
\bibitem{Yuksel:2008cu}
  H.~Yuksel, M.~D.~Kistler, J.~F.~Beacom and A.~M.~Hopkins,
  Astrophys.\ J.\  {\bf 683}, L5 (2008)
  [arXiv:0804.4008 [astro-ph]].

\bibitem{Chakraborty:2008zp}
  S.~Chakraborty, S.~Choubey, B.~Dasgupta and K.~Kar,
  JCAP {\bf 0809}, 013 (2008)
  [arXiv:0805.3131 [hep-ph]].

\bibitem{Haxton:1987bf}
  W.~C.~Haxton and C.~W.~Johnson,
  Nature {\bf 333}, 325 (1988).

\bibitem{Lazauskas:2009yh}
  R.~Lazauskas, C.~Lunardini and C.~Volpe,
  JCAP {\bf 0904}, 029 (2009)
  [arXiv:0901.0581 [astro-ph.SR]].


\bibitem{Botella:1986wy}
  F.~J.~Botella, C.~S.~Lim and W.~J.~Marciano,
  Phys.\ Rev.\  D {\bf 35} (1987) 896.

\bibitem{Gava:2009gt}
  J.~Gava and C.~C.~Jean-Louis,
  arXiv:0907.3947 [hep-ph].

\bibitem{Gava:2008rp}
  J.~Gava and C.~Volpe,
  Phys.\ Rev.\  D {\bf 78}, 083007 (2008)
  [arXiv:0807.3418 [astro-ph]].

\bibitem{Arcones:2006uq}
  A.~Arcones, H.~T.~Janka and L.~Scheck,
  arXiv:astro-ph/0612582.

\bibitem{Baldry:2003xi}
  I.~K.~Baldry and K.~Glazebrook,
  Astrophys.\ J.\  {\bf 593}, 258 (2003)
  [arXiv:astro-ph/0304423].
  
\bibitem{Amsler:2008zzb}
  C.~Amsler {\it et al.}  [Particle Data Group],
  Phys.\ Lett.\  B {\bf 667}, 1 (2008).

\bibitem{Balantekin:2007es}
  A.~B.~Balantekin, J.~Gava and C.~Volpe,
  Phys.\ Lett.\  B {\bf 662}, 396 (2008)
  [arXiv:0710.3112 [astro-ph]].
  
\bibitem{Kneller:2009vd}
  J.~P.~Kneller and G.~C.~McLaughlin,
  Phys.\ Rev.\  D {\bf 80}, 053002 (2009)
  [arXiv:0904.3823 [hep-ph]].

\bibitem{Dighe:1999bi}
  A.~S.~Dighe and A.~Y.~Smirnov,
  Phys.\ Rev.\  D {\bf 62}, 033007 (2000)
  [arXiv:hep-ph/9907423].

\bibitem{Ando:2002zj}
  S.~Ando and K.~Sato,
  Phys.\ Lett.\  B {\bf 559}, 113 (2003)
  [arXiv:astro-ph/0210502].



\bibitem{Lunardini:2007vn}
  C.~Lunardini, B.~Muller and H.~T.~Janka,
  Phys.\ Rev.\  D {\bf 78}, 023016 (2008)
  [arXiv:0712.3000 [astro-ph]].


\bibitem{Fogli:2006xy}
  G.~L.~Fogli, E.~Lisi, A.~Mirizzi and D.~Montanino,
  JCAP {\bf 0606}, 012 (2006)
  [arXiv:hep-ph/0603033].

\bibitem{Friedland:2006ta}
  A.~Friedland and A.~Gruzinov,
  arXiv:astro-ph/0607244.

\bibitem{Duan:2009cd}
  H.~Duan and J.~P.~Kneller,
  J.\ Phys.\ G {\bf 36}, 113201 (2009)
  [arXiv:0904.0974 [astro-ph.HE]].

\bibitem{Dasgupta:2007ws}
  B.~Dasgupta and A.~Dighe,
  Phys.\ Rev.\  D {\bf 77}, 113002 (2008)
  [arXiv:0712.3798 [hep-ph]].

\end{thebibliography}
\end{document}